\documentclass{article}
\usepackage{spconf,amsmath,graphicx}

\usepackage{amssymb}
\usepackage{url}
\usepackage{amsthm}
\usepackage{multirow}
\usepackage{bbm}
\usepackage{caption}
\usepackage{mathtools}
\usepackage{array}
\usepackage{algorithm}
\usepackage{algpseudocode}
\usepackage{amsmath}
\usepackage{verbatim}

\usepackage{graphicx} 
\usepackage{multicol}
\usepackage{epstopdf}
\usepackage{cite}
\usepackage{float}

\usepackage{enumitem}
\usepackage{array}
\usepackage{lipsum}
\usepackage{fixltx2e}
\usepackage{amsfonts}
\usepackage{algpseudocode}%
\usepackage[numbers,square, comma, sort & compress]{natbib}
\usepackage[caption=false,font=normalsize,labelfont=sf,textfont=sf]{subfig}

\usepackage{amsmath}

\newtheorem{thm}{Theorem}
\newtheorem{lemma}{Lemma}

\usepackage[cmintegrals]{newtxmath}

\begin{document}
\title{Distributed Algorithm for Dynamic Cognitive 
Ad-Hoc Networks}

\name{\normalsize{Rohit Kumar$^{\dagger}$, Shaswat Satapathy$^{\star}$, Shivani Singh$^{\star}$, and Sumit J. Darak$^{\dagger}$ }}
\address{\normalsize$^{\star}$CSE Department, IIIT-Bhubaneswar, India,\\ \normalsize$^{\dagger}$ECE Department, IIIT-Delhi, India}


\maketitle

\begin{abstract}
Cognitive ad-hoc networks allow users to access an unlicensed/shared spectrum without the need for any coordination via a central controller and are being envisioned for futuristic ultra-dense wireless networks. The ad-hoc nature of networks require each user to learn and regularly update various network parameters such as channel quality and the number of users, and use learned information to improve the spectrum utilization and minimize collisions. For such a learning and coordination task, we propose a distributed algorithm based on a multi-player multi-armed bandit approach and novel signaling scheme. The proposed algorithm does not need prior knowledge of network parameters (users, channels) and its ability to detect as well as adapt to the changes in the network parameters thereby making it suitable for static as well as dynamic networks. The theoretical analysis and extensive simulation results validate the superiority of the proposed algorithm over existing state-of-the-art algorithms.	
\end{abstract}
\begin{keywords}
	Mutli-player multi-armed bandit, Cognitive ad-hoc networks, Dynamic network, Change detection.
\end{keywords}

\vspace{-0.25cm}
\section{Introduction}

Cognitive ad-hoc networks allow users to access a shared/unlicensed spectrum without the need of any coordination via central controller or control channels \cite{ca1,ca2}. They are being envisioned for futuristic ultra-dense wireless communication networks such as the Internet of Things (IoT) that can offer very high peak rates but low average traffic per user. The ad-hoc nature makes coordination challenging as users not only have to learn and regularly update network parameters such as channel quality and the number of users but also need to use learned parameters to improve the spectrum utilization (i.e., throughput) and minimize collisions. \par

Various distributed algorithms have been proposed to facilitate learning and coordination tasks in static networks where the network parameters do not change with time \cite{prand,MCTOPM,wcnc,wiopt,quek,gai2,zhandi}. The musical chair (MC) based MCTopM algorithm in \cite{MCTOPM} is the current state-of-the-art algorithm for static networks but assumes prior knowledge of the number of users, $N$ which is not practical for ad-hoc networks. When $N$ is unknown, 
the secondary user coordination with fairness (SCF) algorithm in \cite{wcnc} is the state-of-the-art algorithm. The drawback of the SCF algorithm is that it needs prior knowledge of the minimum difference between channel statistics, $\Delta$. In this paper, we consider a more challenging dynamic network where the channel statistics may change with time, as well as the users, can enter
or leave the network any time without prior agreement. To the best of our knowledge, \cite{MEGA,MC} are the only algorithms that consider the dynamic network scenario. The dynamic MC (DMC) algorithm in \cite{MC} has shown to outperform \cite{MEGA}. The trekking based algorithms in \cite{wiopt} can adapt to changing $N$ but assumes stationary channels. The dynamic MC (DMC) algorithm follows a randomized hopping (RH) based \textit{epoch} approach which allows it to adapt to unknown and changing $N$ as well as channel statistics \cite{MC}. But, DMC also needs knowledge of $\Delta$ and it fails when the number of users is more than the number of channels. The RH phase in DMC forces users to select channels uniformly at random which leads to poor throughput due to a large number of collisions and frequent selection of sub-optimal channels. In \cite{CP1,CP2}, MAB algorithms for dynamic channel cases have been proposed but their feasibility for multi-player MAB has not been discussed yet. The design of an algorithm which does not need prior knowledge of $N$, $\Delta$ and can adapt to the need of dynamic networks is the focus of this work.\par

 Proposed, Estimate-Explore-Exploit-Detect-repeat E$^3$DR algorithm enables users to \textit{estimate} the number of active users via novel signaling scheme, learn channel statistics and exploit optimum channels\footnote{For a network with $N$ users, we arrange the channels in the decreasing order of their average throughput. Then, the set of first $N$ channels are referred to as optimum channels.} via \textit{explore-exploit} based MAB algorithm and adapt to the changes in these parameters via change \textit{detection} approach.
\vspace{-0.25cm}

\section{Network Model}\label{network_model}

We consider a cognitive ad-hoc network where users can transmit over $K$ channels. Similar to \cite{wcnc,wiopt,MCTOPM,MC}, the throughput obtained by user when it transmits over the channel, $k$, $k\in[K]$, is sampled independently from some distribution on [0,1] with mean $\mu_{k}$. We consider a dynamic environment where channel statistics may change after certain unknown intervals. For instance, we can divide the time horizon into $B$ blocks such that the channel statistics remain the same in a block $b$, $b\in[B]$ but may change from one block to another. Thus, channel statistics are denoted as $\mu_{k,b}$. Each user can transmit only once in a time slot and when multiple users transmit simultaneously on the same channel, a collision occurs leading to zero throughput. 
The usefulness of the distributed algorithm is validated using the metric regret which is the difference between expected optimal throughput and run-time average throughput. Mathematically,\vspace{-0.3cm}
	 \begin{align} \label{eq:regret} 
	 Regret &= R_{op}-\sum\limits_{b=1}^{B}\sum\limits_{t=1}^{T_b}\sum\limits_{n=1}^{N} \mu_{A_{t,b}^n}(1- E\left[C_{A_{t,b}^n}\right]).	 
	 \end{align}
	where $R_{op}$ is the maximum mean total throughput achievable for users. It is achieved when users are orthogonalized (i.e. no collision) and locked on the $N$ optimum channels at the start of every block. 
	$A_{t,b}^n$ denotes the channel selected by $n^{th}$ user at time $t$ in block $b$. $\mu_{A_{t,b}^n}$  and $C_{A_{t,b}^n}$ denote the expected throughput and collision indicator on channel  $A_{t,b}^n$, respectively. If there is collision, collision indicator is set to $1$, otherwise it is $0$. Our goal is to develop distributed algorithm that minimizes regret (i.e. throughput loss).
	 \vspace{-0.2cm}

\vspace{-0.25cm}
\section{Proposed E$^3$DR Algorithm}\label{SN}
\vspace{-0.25cm}
The proposed E$^3$DR algorithm consists of four phases, namely Orthogonalization, Estimate, Explore-Exploit and Detect phase that runs sequentially and repeats at the regular interval. For clarity of notations, we omit subscripts $b$ and $n$.
\vspace{-0.35cm}
\subsection{ Orthogonalization (OR) Phase}
\vspace{-0.25cm}
The \textit{OR} phase of duration $T_O$ is similar to \cite{wcnc,wiopt}. Each user selects the channel uniformly at random and transmits over it. If the transmission is collision-free, i.e., successful, the user locks himself on that channel till the end of the OR phase. Let us denote this channel as $A$. In case of collision, an unlocked user repeats the same process in subsequent time slots until it gets locked. When $N>K$ i.e., the number of active users is more than channels, some users may not get locked and hence, they must back-off and re-enter in next OR phase. 
\begin{lemma}
	\label{lma:RanHop_Static}
	Let $\delta \in (0,1)$. If OR phase runs for $T_{O}:=\left \lceil \frac{\log(\delta_1/K)}{ \log\left(1-1/4K\right)} \right \rceil $ number of time slots, then all the users will orthogonalize with probability at least $1-\delta$.  
\end{lemma} \vspace{-0.15cm}
\noindent \textbf{Proof:} The proof of this lemma is simple and derived by extending the
proof of Lemma 1 in \cite{wiopt} for unlicensed spectrum.
	\vspace{-0.45cm}
\begin{figure}[!b]
	\vspace{-0.45cm}
	\centering
	\includegraphics[scale=0.7]{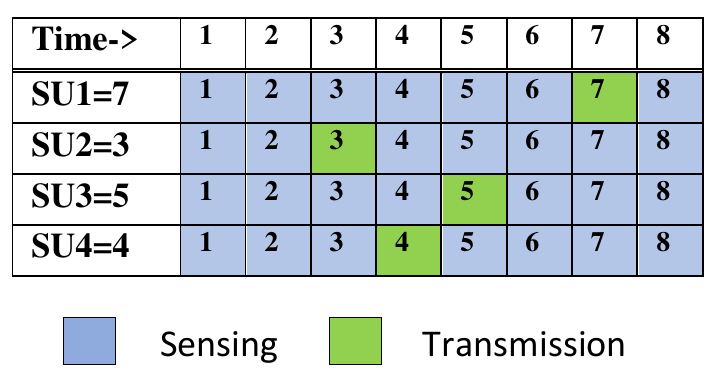}%
	\vspace{-0.25cm}
	\label{static}
	\caption{{\footnotesize Illustrative example of signalling scheme with $K=8$ and $N=4$.}}
	\vspace{-0.15cm}
\end{figure}
\subsection{Estimate Phase}
\vspace{-0.25cm}
The \textit{estimate} phase given in Subroutine 1 allows the user to estimate the number of active users using the novel signaling scheme shown in Fig.1. 
%
Note that all users are locked on distinct channels at the end of \textit{OR} phase. In Subroutine 1, users with sensing capability i.e., ability to detect the presence of other users without any transmission, sequentially sense the channel as per channel index, i.e., the user senses the first channel in the first time slot, second channel in the second time slot and so on. The exception is the $A^{th}$ time slot where the user transmits over the $A^{th}$ channel instead of sensing as shown in Fig. 1. Then, by counting the number of sensed transmissions, the user estimates the number of locked users, $N$. Thus, the duration of this phase is $T_{Est}=K$ time slots.\footnote{Users without sensing capability perform the estimation in $K$ frames each consisting of $K$ time slots i. e., total $K^2$ time slots. A user on channel $A$ at the end of the OR phase transmits on the same channel in all the frames except $A^{th}$ frame. In the $A^{th}$ frame, user sequentially transmit on each channel and estimates $N$ based on the number of collisions incurred.} 


\begin{algorithm}[!t]
	\caption*{\textbf{Subroutine 1:} Estimate Phase}
	\begin{algorithmic}[1]
		\State Input: $K, A$
		\State Set $N=1$ 
	    \For{$t=1 \dots K$}
	    \If {$A == t$}
	    \State Transmit on the channel $A$
	    \Else
	    \State Sense channel index $t$. If busy, $N=N+1$
		\EndIf
	    \EndFor	
	\end{algorithmic}
\end{algorithm}
\vspace{-0.25cm}
\begin{lemma}
	\label{lma:est}
	For all $\delta \in (0,1)$, with probability at least
$1 - \delta$ , all the users will have correct estimation of number of users if the estimate phase is run for $T_{Est}=K$ time slots.
\end{lemma} \vspace{-0.25cm}
\noindent \textbf{Proof:} In each time slot, all other SUs will be able to sense the user who is transmitting in that time slot. Thus, in $K$ time slots, all the SUs will be able to sense and thus, estimate all the $T_{Est}=K$ (maximum possible) users in the network.

\vspace{-0.25cm}
\subsection{Explore-Exploit Phase}
\vspace{-0.25cm}
After estimation of the number of users, each user needs to learn channel statistics and exploit optimum $N$ channels. This results in an exploration-exploitation trade-off since users need to select all channels sufficient number of times to learn their statistics and at the same time, the selection of sub-optimum channels should be minimized. Furthermore, in multi-user networks, collisions should be as small as possible.

Though the sequential hopping (SH) approach in the SCF algorithm \cite{wcnc,wiopt} enables learning of channel statistics without incurring any collision, it needs knowledge of $\Delta$. In the proposed \textit{Explore-Exploit} phase, each user employs MCTopM algorithm in \cite{MCTOPM}. Since users have estimated $N$, they can now use state-of-the art MCTopM MAB algorithm. It is based on a multi-user MAB approach using the upper confidence bound (UCB) algorithm and it guarantees orthogonalization of users over optimum $N$ channels with high probability. However, the MCTopM algorithm incurs significant regret in the beginning due to exploration (small $t$) and the probability of exploration decreases as $t$ increases. In the proposed approach, we reset the MCTopM algorithm only when channel statistics changes otherwise MCTopM continues from the previous epoch leading to less exploration and lower regret. In each epoch, McTopM is run for $T_M=(2000)$ time slots.

\vspace{-0.25cm}
\subsection{Detect Phase}
\vspace{-0.25cm}
The \textit{detect} phase given in Subroutine 2 enables users to detect changes in the channel statistics. It takes an index of the channel $A$, say $I$ selected in OR phase and the estimated channel statistics, $\hat{\mu}$ in the explore-exploit phase as input. For a single-user network, the user needs to sense all channels sequentially a sufficient number of times, say $T_D$. In the end, the statistics learned during this period are compared with previously learned statistics. If the difference between these two statistics is greater than a threshold, $\psi=\frac{\Delta}{2}$, for at least one channel, it is assumed that channel statistics have been changed. For a multi-user case, this approach needs $KT_D$ time slots for each user. In the proposed algorithm, we exploit multiple users in the networks to reduce the duration of \textit{detect} phase. Each user senses $\lceil \frac{K}{N} \rceil$ channels and to avoid multiple users sensing the same channel, the channel index is based on the user index, $I$. After sensing one channel, all users inform other users via a signaling scheme similar to \textit{estimate} phase. Thus, the duration of \textit{detect phase} is reduced to $T_{DD}=T_D\lceil \frac{K}{N} \rceil$. To avoid frequent resetting, we choose $\psi>0.05$.

\begin{algorithm}[!h]
	\caption*{\textbf{Subroutine 2:} Detect Phase}
	\begin{algorithmic}[1]
		\State Input: $K,N,I, \hat{\mu}$
		\State $A$=OR$(K)$
		\For {$v=1 \dots \lceil \frac{K}{N} \rceil$}
		\For {$t=1 \dots T_{D}$}
		\State Transmit on channel with index, a = $(I-1)\lceil \frac{K}{N} \rceil+v$
		\State Update $X_{a} = X_{a} + r_{t,a}$ and $Y_{a}=Y_{a}+1$
		\EndFor
		\State $\tilde{\mu}_{a} = \frac{X_{a}}{Y_{a}}$
		\For{$t=1 \dots N$}
	   \If {$t==I$}
	    \If{ ($\mid\tilde{\mu}_{a} - \hat{\mu}_{a}\mid \geq \psi)$}
		\State Set $D=1$ and transmit on the channel $a$
		\EndIf
	    \Else
	    \State Sense channel with index $(t-1)\lceil \frac{K}{N} \rceil+v$. 
	    \State Set $D=1$ if channel is busy.
		\EndIf
	    \EndFor
	    \EndFor	
	\end{algorithmic}
\end{algorithm}

\begin{lemma}
	\label{lma:detect}
	If each user selects the channel consecutively for $T_{D}=\frac{2}{\epsilon^2} \cdot \ln \left ( \frac{2}{\delta} \right)$ number of time slots and does not incur any collision, then with probability at least $1-\delta$, the user will have the $\epsilon-$ correct detection of change in the channel statistics.
\end{lemma} 
\noindent \textbf{Proof:} An user has $\epsilon-$ correct detection of change in the channel statistics if $\mid\tilde{\mu}_{k} - \hat{\mu}_{k}\mid \leq \frac{\epsilon}{2} \forall k \in 1 \cdots K$.  We upper bound the probability that the user has $\epsilon-$ correct detection of the change in the channel statistics given the user selects that channel $T_{D}$ number of times. 
\begin{align*}
Pr\left ( \exists k \in 1 \cdots K \ \ s.t | \mid\tilde{\mu}_{k} - \hat{\mu}_{k}\mid \leq \frac{\epsilon}{2} \right | T_D) \\
\leq \left(1-2 \cdot \exp \left(- 2 \cdot T_D \cdot \frac{\epsilon^2}{4} \right)\right) \mbox{(Using Hoeffding's Inequality)}
\end{align*}

In order for this to be $> 1- \delta$, we set
\[2 \cdot \exp \left(\frac{-T_{D} \cdot \epsilon^2}{2} \right) < \delta \newline
\implies  T_{D} >  \frac{2}{\epsilon^2}  \ln \left ( \frac{2}{\delta}   \right).
\]

\begin{thm}
	\label{thm:regret_Static}
For all $\delta \in (0,1)$, with probability  $\geq 1-\delta$,  the expected regret of the E$^3$DR algorithm over $T$ rounds for the network consisting of $N$ users and $K$ channels is at most: $R_T <  R_{SE} \bigg(\frac{2 \cdot T}{T_{EP}}-e \bigg) + l(T_{EP}-T_{Est})$. 
\end{thm} 

\noindent \textbf{Proof:} Assume $T_{EP}$ be the epoch length. The number of epochs is at most $\lceil \frac{T}{T_{EP}}\rceil$. Let $N_m$ be the number of SUs who enter in the network at the start of the horizon, $e$ and $l$ be the total number of SUs entering and leaving the network. 

\noindent \textbf{Regret due to the OR, Estimate, Explore-Exploit and Detect phase:} Let $N_m \leq K$ be the number of users at the start of the epoch. Regret incurred by the users during the OR and estimate phase is upper bounded by $N \cdot T_{O}$ and $N \cdot (K-1)$. The regret incurred in the Explore-Exploit phase is upper bounded by $\mathcal{O}(\log T)$ \cite{MCTOPM}. Whereas the regret incurred during the detect phase is given by $\lceil \frac{K}{N} \rceil (T_{D}-1)$. Note that users will not incur regret for $\lceil \frac{K}{N} \rceil$ fraction of detect phase as they will select any of the optimal channels. Thus, the contribution to the regret in that time slot is zero. Since an Explore-Exploit and detect phase may be run for more than once in an epoch depending upon the epoch size, thus expected regret of the E$^3$DR algorithm incurred in a single epoch is given by: 
\begin{equation*}
R_{SE}=N_m \cdot (T_{O}+(K-1)) + x \cdot \big\{\mathcal{O}(\log T) + \lceil \frac{K}{N_m} \rceil (T_{D}-1)\big\}
\end{equation*}

where $x=\lceil \frac{(T_{EP}-T_O-T_{Est})}{(T_{EE}+T_{DD}+K} \rceil$.

\noindent \textbf{Regret due to the entering users:} A new user
may enter in the network earliest in the start of the second epoch. Thereafter, it will incur regret similar to the users who are in the network from the start of the horizon.

\noindent  \textbf{Regret due to the leaving users:} Recall that the user can leave any time except during the estimate phase. If user leaves, one of the optimal channels may remain unused and the regret is incurred till the end of that epoch. Hence, if $l_i$ user leave in an epoch, it add at most $l_i(T_{EP}-T_{Est})$ regret.

Let $e=\sum_{i=1}^{T/T_{EP}} e_i$ and $l=\sum_{i=1}^{T/T_{EP}} l_i$ denote the total number of entering and leaving SUs across all epochs. Combining the regret over all the epochs, we get the total expected regret of the E$^3$DR algorithm as:

 \begin{equation*}
  R_T <  R_{SE} \bigg(\frac{2 \cdot T}{T_{EP}}-e \bigg) + l(T_{EP}-T_{Est}).
\end{equation*}

\begin{thm}
	\label{thm:Collsions_Static}
For any given $\delta \in (0,1)$, with probability  $\geq 1-\delta$,  the expected number of collisions incurred by the E$^3$DR algorithm in $T$ rounds and the network consisting of $N$ users and $K$ channels is at most $\mathcal{O}(\log T)$. 
\end{thm} \vspace{-0.25cm}
\noindent \textbf{Proof:} Number of collisions during the OR phase is upper bounded by: $N\cdot T_{O}$. Whereas, the number of collisions in the Explore-Exploit phase is upper bounded by $\mathcal{O}(\log T)$ \cite{MCTOPM}. Other phases do not have any collisions. Thus, the the expected number of collisions incurred by the E$^3$DR over all the epoch is upper bounded by: $
\frac{T}{T_{EP}} \big( N \cdot T_{O} + \mathcal{O}(\log T) \big).
$
\vspace{-0.25cm}
\section{Simulation Results}\label{sim_res}

\vspace{-0.25cm}
We compare the performance of state-of-the-art MCTopM \cite{MCTOPM} (known $N$), SCF \cite{wcnc} and TSN \cite{wiopt} algorithms (unknown $N$ but known $\Delta$) and E$^3$DR algorithm (unknown $N$ and $\Delta$) in terms of regret (i.e. throughput loss). These algorithms have been selected as they have shown to outperform DMC \cite{MC} and algorithms such as \cite{prand,gai2,MEGA,zhandi} and hence, we don't include the rest to maintain the clarity of plots. All the results presented here are obtained after averaging over 50 experiments and each experiment consists of the horizon of $10^5$ time slots. The channel statistics are chosen randomly in each experiment. Entering and leaving users and change points in the channel statistics are shown using red, yellow and blue dashed lines, respectively.
We begin with the static network where number of users are fixed throughout the horizon and channel statistics are stationary. We consider $K=10$ and $N=4$. As shown in Fig.~2, the MCTopM has lowest regret as it has prior knowledge of $N$. The proposed E$^3$DR algorithm significantly outperforms SCF and TSN algorithms. Note that after initial \textit{OR} and \textit{Estimate} phase, E$^3$DR and MCTopM algorithms incur identical regret per slot. Higher regret in the beginning is the penalty due to unknown $N$ which cannot be avoided. 

\begin{figure}[!h]
	\centering
	\includegraphics[scale=0.3]{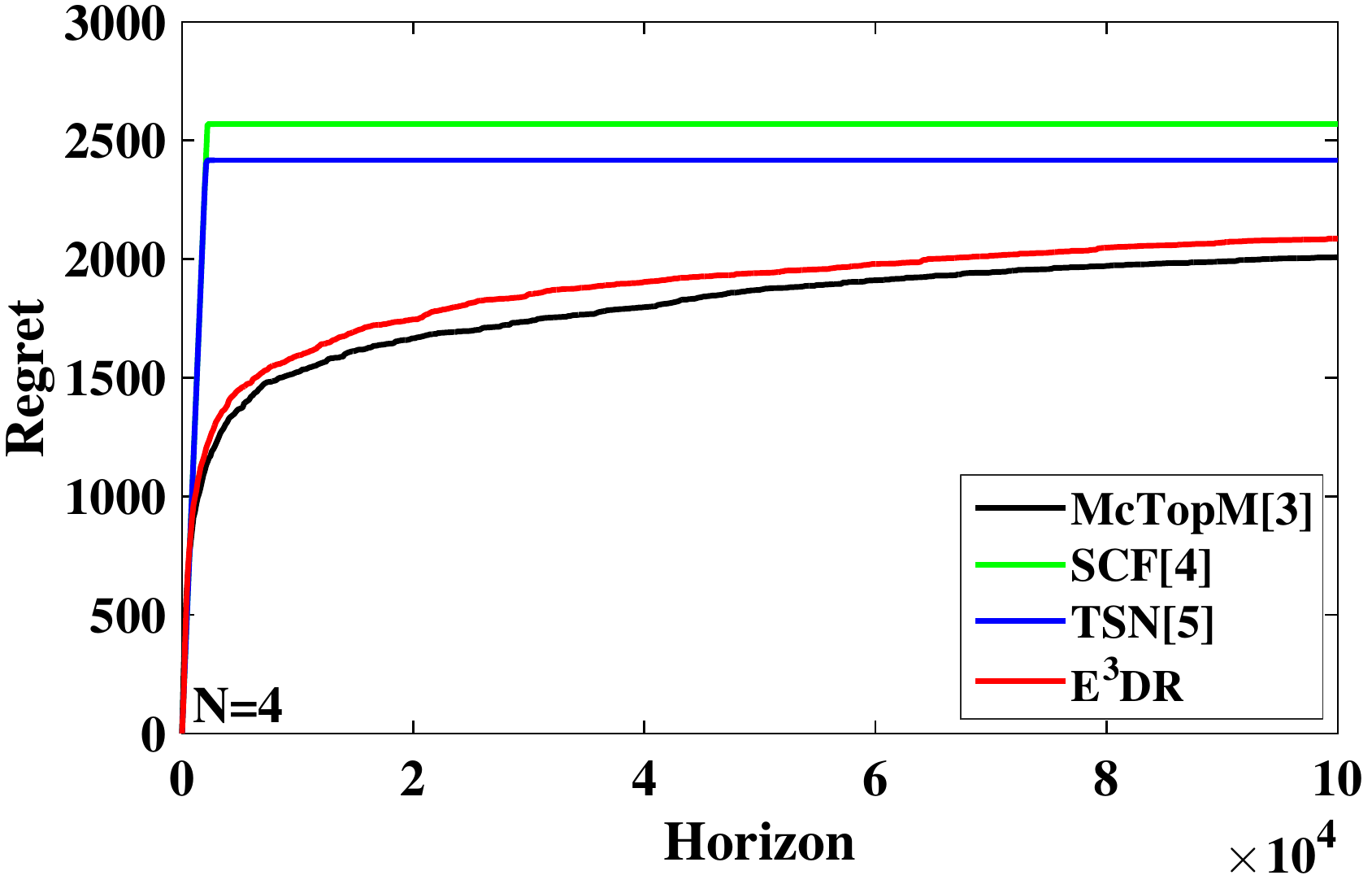}%
	\label{sd}
	\vspace{-0.2cm}
	\caption{{\footnotesize The comparison for average regret of various algorithms for Static network with $N=4$. Lower is better.}}
	\vspace{-0.25cm}
\end{figure}

\begin{figure}[!h]
	\vspace{-0.3cm}
	\centering
	\subfloat[]{\includegraphics[scale=0.25]{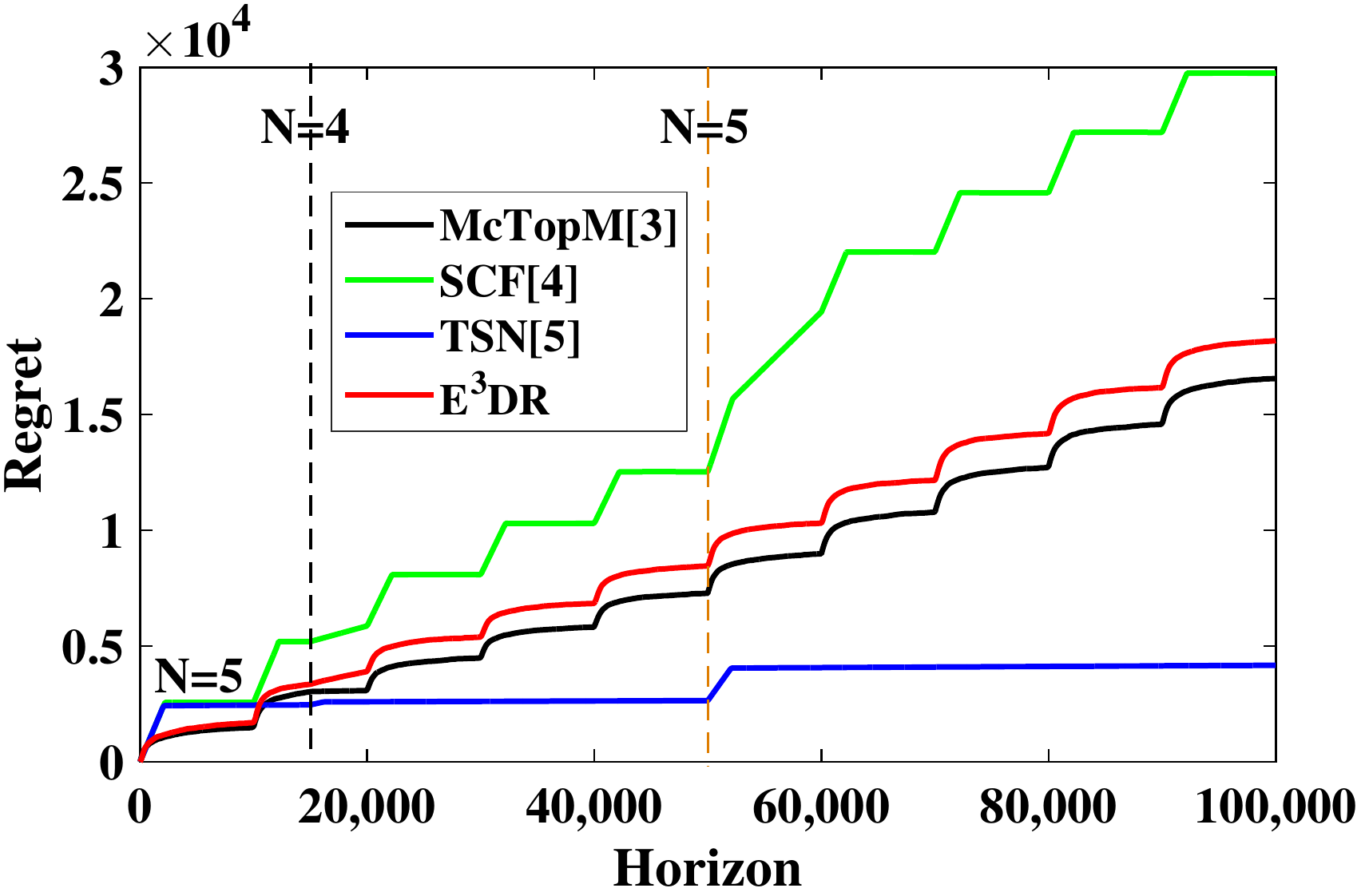}
		\label{dynamic_users}}
	\subfloat[]{\includegraphics[scale=0.25]{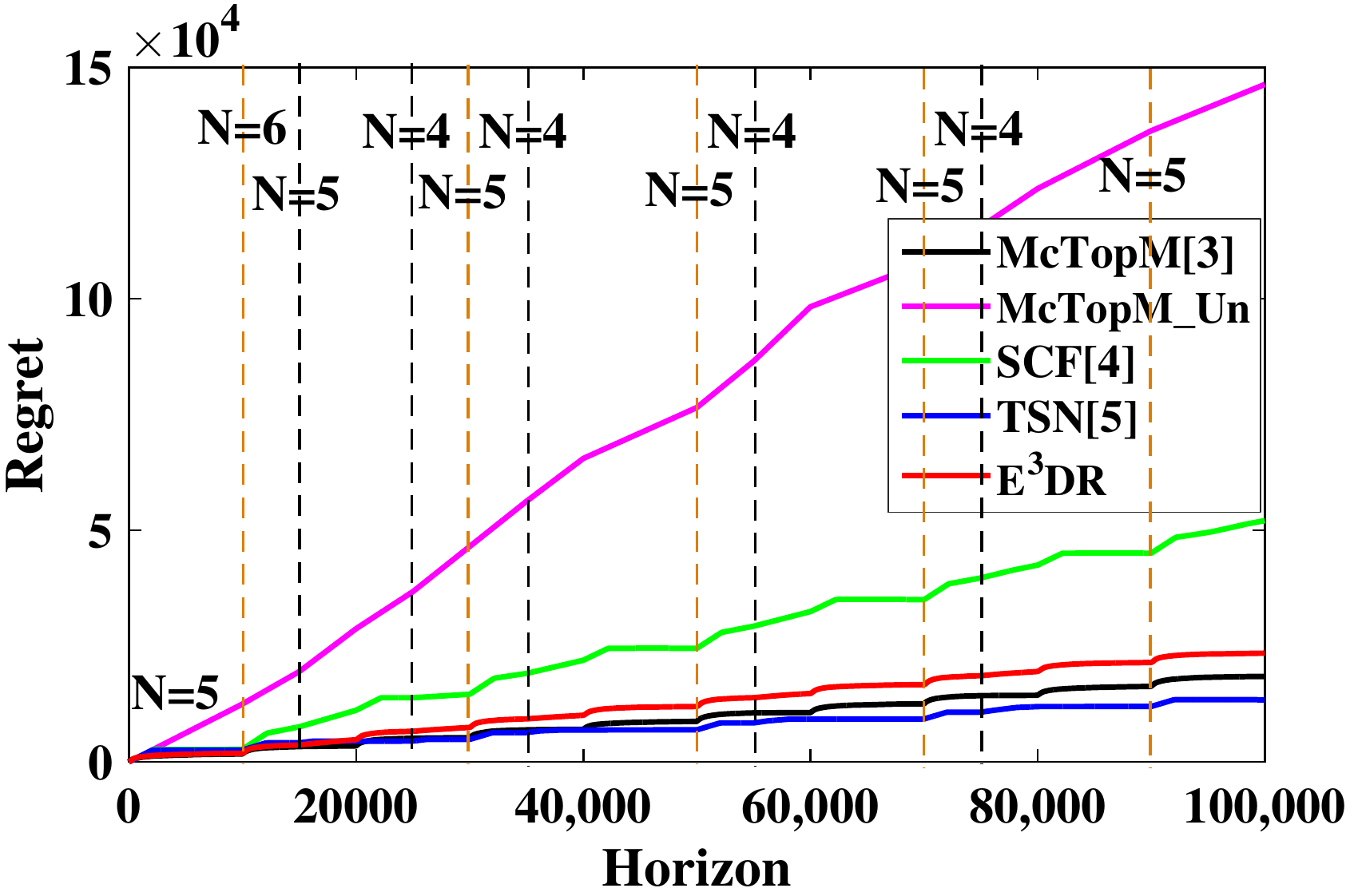}
		\label{dynamic_users1}}\\\vspace{-0.45cm}
	\subfloat[]{\includegraphics[scale=0.25]{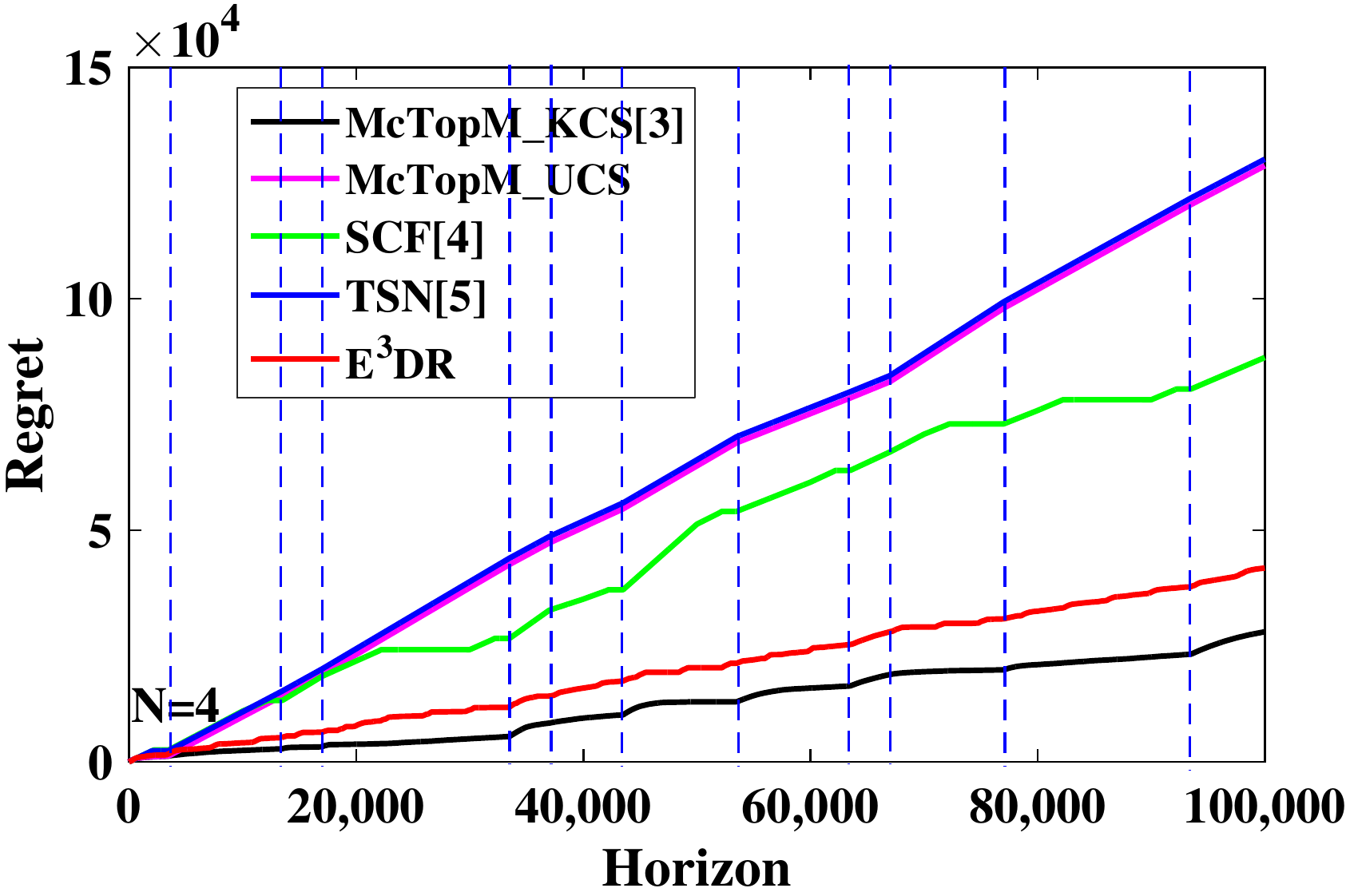}
		\label{dynamic_channel}}
	\subfloat[]{\includegraphics[scale=0.25]{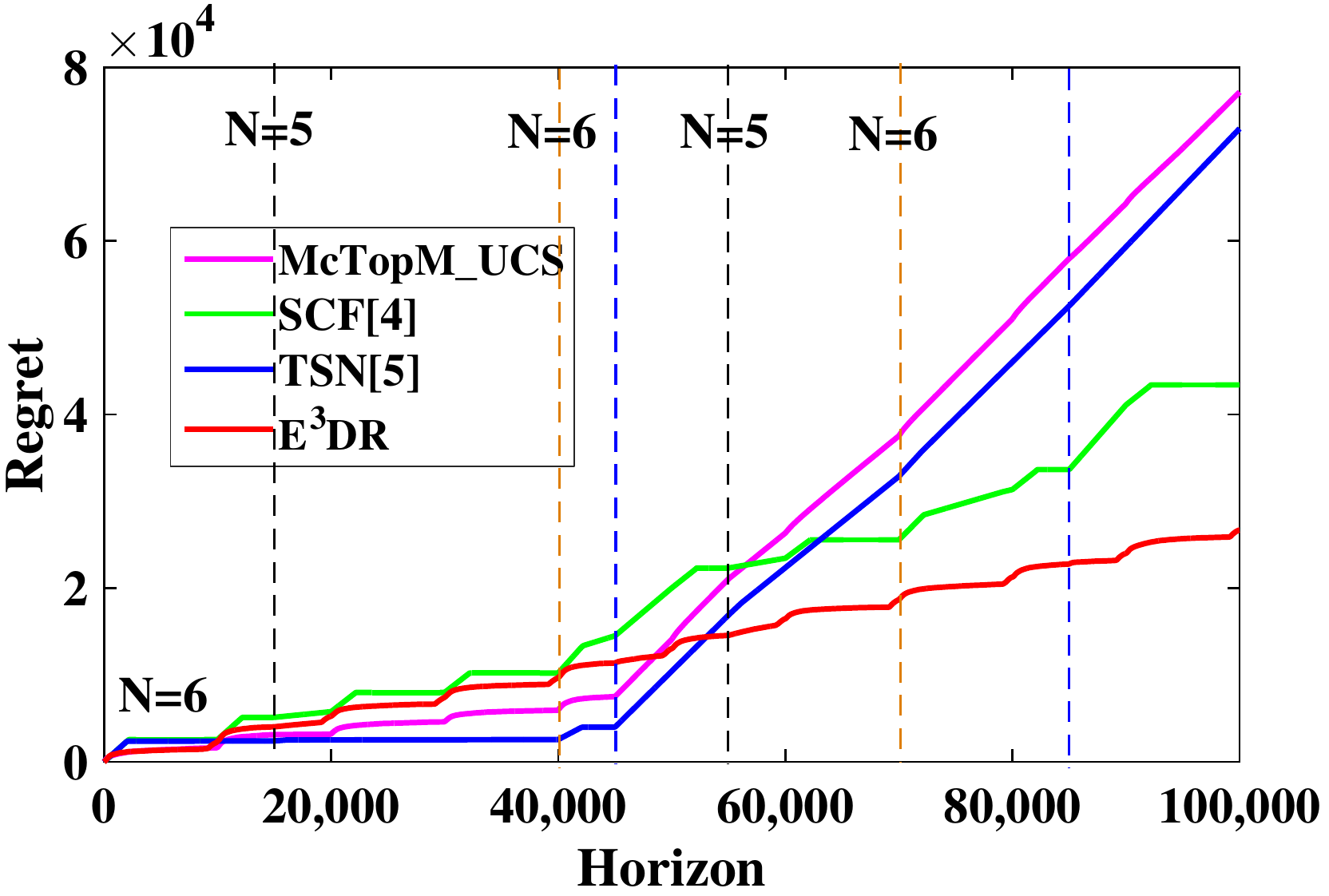}
		\label{dyn_ch_user}}
	\vspace{-0.45cm}
	\caption{{\footnotesize The comparison for average regret of various algorithms for Dynamic network (a) Case 1 with fewer entry/exits, (b) Case 1 with frequent entry and exits, (c)  Case 2, and (d) Case 3. Lower is better.}}
	\vspace{-0.35cm}
	\label{reg2u}
\end{figure}

Next, we consider the dynamic networks with three cases: 1) Stationary channels and dynamic users where users can enter or leave, 2) Quasi-stationary channels i. e., changing channel conditions and a fixed number of users, and 3) Quasi-stationary channels and dynamic users. 

For Case 1, there are five users in the beginning and the leaving user is chosen randomly from the set of active users. As we move from Fig.~\ref{dynamic_users} to Fig.~\ref{dynamic_users1}, the difference between MCTopM and E$^3$DR algorithms decreases in spite of former having prior knowledge of changing $N$. 
This is because MCTopM needs a new user to learn channel statistics and hence, there is an additional regret for every new user. It is also evident from Fig.~\ref{dynamic_users1} that McTopM performance degrades significantly when it does not know $N$ (See McTopM$\_{\mbox{Un}}$). AS expected, the TSN algorithm offers the lowest regret for fixed channel statistics but it performs poorly for quasi-stationary channels as discussed next. 

For Case 2, regret plots are shown in Fig.~\ref{dynamic_channel} where channel statistics change frequently. New statistics are chosen randomly. It can be observed that the epoch approach based SCF algorithm incurs regret at the regular interval while E$^3$DR algorithm offers lower regret due to a novel change detection approach. Although the  MCTopM with prior information about the change in channel statistics (refereed as McTopM$\_{\mbox{KCS}}$), it performs slightly better than E$^3$DR algorithm. However, its performance degrades drastically when such information is not available (See McTopM$\_{\mbox{UCS}}$). Similarly, the TSN incurs significant regret due to the failure to adapt to the changes in channel statistics. Finally, in Case 3, we consider a challenging scenario by allowing the channel statistics and the number of users to change at any time. It can be observed that our algorithm offers the lowest regret.

\vspace{-0.25cm}
\section{Conclusions and Future Works} \label{conclusion}
\vspace{-0.25cm}
In this paper, the Estimate-Explore-Exploit-Detect-repeat (E$^3$DR) algorithm for dynamic cognitive ad-hoc network is proposed. It allow users to learn to coordinate and adapt without direct communication,  does not need prior knowledge of channel statistics, $\Delta$, $N$ and enable new users to back-off when $N>K$. Simulation results show that the E$^3$DR algorithm offers superior performance over existing state-of-the-art algorithms. Future work includes the extension of proposed algorithm for dynamic networks where channel statistics are different at each user such as \cite{leshem,jsac1}.

\end{document}